\input harvmac
\noblackbox

\input epsf
\def\figin{\epsfcheck\figin}\def\figins{\epsfcheck\figins}
\def\epsfcheck{\ifx\epsfbox\UnDeFiNeD
\message{(NO epsf.tex, FIGURES WILL BE IGNORED)}
\gdef\figin##1{\vskip2in}\gdef\figins##1{\hskip.5in}
\else\message{(FIGURES WILL BE INCLUDED)}%
\gdef\figin##1{##1}\gdef\figins##1{##1}\fi}
\def\DefWarn#1{}
\def\figinsert{\goodbreak\midinsert}
\def\ifig#1#2#3{\DefWarn#1\xdef#1{fig.~\the\figno}
\writedef{#1\leftbracket fig.\noexpand~\the\figno}%
\figinsert\figin{\centerline{#3}}\medskip\centerline{\vbox{\baselineskip12pt
\advance\hsize by -1truein\noindent\footnotefont{\bf Fig.~\the\figno:} #2}}
\bigskip\endinsert\global\advance\figno by1}

\def\d{{\partial}}

\lref\natidia{D-E. Diaconescu and N. Seiberg, {\it The Coulomb Branch of 
(4,4) Supersymmetric Field Theories in Two Dimensions}, IAS-97-84, RU-97-62, 
hep-th/9707158.}

\lref\coleman{S. Coleman}

\lref\hw{A. Hanany and E. Witten, {\it Type IIB Superstrings, BPS Monopoles, 
And Three-Dimensional Gauge Dynamics}, IASSNS-HEP-96/121, hep-th/9611230.}

\lref\amihori{A. Hanany and K. Hori, {\it Branes and N=2 Theories in Two Dimensions}, IASSNS-HEP-97/81, UCB-PTH-97/39, LBNL-40571, hep-th/9707192.}

\lref\bending{E. Witten, {\it Solutions Of Four-Dimensional Field 
Theories Via M Theory}, hep-th/9703166.}

\lref\phases{E. Witten, {\it Phases of $N=2$ Theories In Two Dimensions}, 
Nucl.Phys. B403 (1993) 159-222, hep-th/9301042.}

\lref\ed{E. Witten, {\it On The Conformal Field Theory Of The Higgs Branch}, 
hep-th/9707093.}

\lref\oogurivafa{H. Ooguri and C. Vafa, {\it Geometry of N=1 Dualities in Four Dimensions}, HUPT-97/A010, UCB-PTH-97/11, LBNL-40032, hep-th/9702180.}

\lref\callan{C. Callan, J. Harvey, and A. Strominger, {\it World Sheet 
Approach to Heterotic Instantons and Solitons}, Nucl.Phys.B359:611-634,1991.}

\lref\ghm{R. Gregory, J. Harvey, and G. Moore, {\it Unwinding strings and 
T-duality of Kaluza-Klein and H-Monopoles}, DTP/97/31, EFI-97-26, YCTP-P15-97, 
hep-th/9708086.}

\lref\mp{D. Morrison and R. Plesser, {\it Towards Mirror Symmetry 
as Duality for Two-Dimensional
Abelian Gauge Theories}, Nucl.Phys.Proc.Suppl. 46 (1996) 177-186, 
hep-th/9508107.}

\lref\ovi{H. Ooguri and C. Vafa, {\it Two-Dimensional Black Hole 
and Singularities of CY Manifolds}, 
Nucl.Phys. B463 (1996) 55-72, hep-th/9511164.}

\lref\qu{F. Quevedo, {\it Duality and Global Symmetries},IFUNAM FT97-7, 
hep-th/9706210.}

\lref\diagomes{D-E. Daiconsecu and J. Gomes, {\it Duality in Matrix 
Theory and Three Dimensional Mirror Symmetry}, RU-97-53, hep-th/9707019.}

\lref\kennati{K. Intriligator and N. Seiberg, {\it Mirror Symmetry in Three 
Dimensional Gauge Theories}, Phys.Lett. B387 (1996) 513, hep-th/9607207.}

\lref\somecomments{E. Witten, {\it Some Comments on String Dynamics}, 
IASSNS-HEP-95-63, hep-th/9507121}

\lref\edINamst{E. Witten, talk at Strings 97 in 
Amsterdam.}

\lref\rv{See for example M. Rocek and E. Verlinde, 
{\it Duality, Quotients, and Currents}, 
Nucl. Phys. B373 (1992) 630-646, hep-th/9110053.}

\lref\bd{M. Berkooz and M. Douglas}

\lref\sethi{S. Sethi and M. Stern}

\lref\schwarz{J. Schwarz}

\lref\fivedbh{D.V.V. {\it Five Dimensional Black Hole}}

\lref\qcdstring{E. Witten, {\it Branes And The Dynamics Of QCD}, 
hep-th/9706109}

\lref\keg{S. Elitzur, A. Giveon, and D. Kutasov, {\it Brane Dynamics 
and N=1 Supersymmetric Gauge Theory}, WIS/97/9, hep-th/9704104.}

\lref\buscher{T.H. Buscher, {\it A Symmetry of the String Background Field 
Equations}, Phys.Lett.194B:59, 1987.}

\lref\dia{D-E. Diaconescu, {\it D-branes, Monopoles and Nahm Equations}, 
hep-th/9608163.}

\lref\variousDIM{E. Witten, {\it String Theory in Various Dimensions}, 
Nucl.Phys. B443 (1995) 85-126, hep-th/9503124.}

\lref\banks{T. Banks, M. Dine, H. Dykstra, W. Fischler, {\it Magnetic Monopole
 Solutions of String Theory}, Phys.Lett.212B:45,1988.}

\lref\ah{Athyah and Hitchin}

\lref\sbr{M. Berkooz, M. Rozali, N. Seiberg, {\it Matrix Description of 
M-theory on $T^4$ and $T^5$}, UTTG-12-97, RU-97-23, hep-th/9704089.}

\lref\newtheories{N. Seiberg, 
{\it New Theories in Six Dimensions and Matrix Description of M-theory on 
$T^5$ and $T^5/Z_2$}, RU-97-42,hep-th/9705221.}

\lref\strominger{A. Strominger, {\it Open p-branes}, 
Phys.Lett. B383 (1996) 44-47, hep-th/9512059.}

\lref\nati{N. Seiberg, {\it IR Dynamics on Branes and Space-Time Geometry},
Phys.Lett. B384 (1996) 81, hep-th/9606017.}

\lref\ednati{N. Seiberg and E. Witten, {\it Gauge Dynamics And 
Compactification To Three Dimensions}, hep-th/9607163.}

\lref\oferami{O. Aharony and A. Hanany, {\it Branes, Superpotentials and Superconformal Fixed Points}, RU-97-25, IASSNS-HEP-97/38, hep-th/9704170.}

\lref\karch{ I. Brunner, A. Karch, {\it Branes and Six Dimensional 
Fixed Points}, HUB-EP-97/27, hep-th/9705022.}

\lref\amigilad{A. Hanany and G. Lifschytz, {\it M(atrix) Theory on $T^6$ 
and a m(atrix) Theory Description of KK Monopoles}, 
IASSNS-HEP-97/93, PUPT-1717, hep-th/9708037.}

\lref\chalmers{G. Chalmers and A. Hanany, {\it Three Dimensional 
Gauge Theories and Monopoles}, Nucl.Phys. B489 (1997) 223-244, hep-th/9608105.}

\lref\townsend{P.K.  Townsend, {\it D-branes from M-branes},  
Phys.Lett. B373 (1996) 68-75,  R/95/59, hep-th/9512062.}

\lref\neqone{N. Seiberg, {\it Electric-Magnetic Duality in 
Supersymmetric Non-Abelian Gauge Theories},  Nucl.Phys. B435 (1995) 129-146,
hep-th/9411149.}

\lref\rocek{ S.J. Gates, Jr., C.M. Hull, and M. Rocek, {\it Twisted Multiplets
and New Supersymmetric Nonlinear Sigma Models.}, Nucl.Phys.B248:157,1984.}

\lref\asp{P. S. Aspinwall, {\it Enhanced Gauge Symmetries and K3 Surfaces}, 
Phys. Lett. B357 (1995) 329, hep-th/9507012.}

\lref\brodieAMI{}
\lref\amiZAPH{}

\Title{\vbox{\rightline{hep-th/9709228}\rightline{PUPT--1725 }
\rightline{}}}
{\vbox{\centerline{Two Dimensional Mirror Symmetry}
\medskip
\centerline{from M-theory}}}
\bigskip
\medskip
\centerline{John H. Brodie}
\smallskip
{\it
\centerline{Department of Physics}
\centerline{Princeton University}
\centerline{Princeton, NJ 08540, USA}}
\centerline{\tt jhbrodie@princeton.edu}
\bigskip
\medskip

\vglue .3cm

\noindent
We construct two dimensional gauge theories with $N= (4,4)$ supersymmetry
from branes of type IIA string theory. 
Quantum effects in the two dimensional gauge theory are analyzed
by embedding 
the IIA brane construction into M-theory. We find that the Coulomb 
branch of one theory and the Higgs branch of a mirror theory 
become equivalent at strong coupling.
A relationship to the 
decoupling limit of the type IIA and IIB 5-branes in Matrix theory is 
shown. T-duality between the ALE metric and the  
wormhole metric of Callan, Harvey, and Strominger 
is discussed from a brane perspective and some puzzles 
regarding string duality resolved. 
We comment on the existence of a quantum Higgs branch in 
two dimensional theories. 
Branes prove to be useful tools in analyzing singular conformal 
field theories.  
\Date{9/97}

\newsec{Introduction}

There has been much progress made recently in understanding field 
theory from brane constructions. In \hw, three dimensional gauge theories 
with 8 real supercharges were studied from type IIB string theory 
using D-branes and Neveu-Schwarz 5-branes. The field theory 
results of \refs{\nati, 
\ednati, \kennati, \chalmers} were obtained from string theory 
and generalized. Specifically, 
the mirror symmetry of three dimensional gauge theories which 
relates hypermultiplets and vector multiplets of two different theories 
 was seen as a result of the S-duality of type IIB. The 
brane construction of three dimensional gauge theories
was generalized to constructions of five dimensional \oferami\ 
and six dimensional \karch\ 
gauge theories with 8 real supercharges. Non-trivial fixed points 
of these gauge theories were derived from string theory 
and associated with tensionless branes. In four dimensions,
brane constructions in type IIA string theory were found to 
be very useful in naturally providing for generalizations of Seiberg-Witten 
type solutions \bending. More recently, a IIB brane construction was
used to study 0+1 dimensional theories in relation to Matrix theory
\amigilad.

In this paper, we examine brane constructions 
of two dimensional gauge theories with 
8 real supercharges. $N_c$ D2-branes are suspended 
between two NS 5-branes with $N_f$ D4 branes intersecting 
the D2-branes in two dimensions. The theory on the 
intersection is then an  
$U(N_c)$ gauge theories with $N_f$ flavors.
The coupling constant of the theory on the D2-brane is 
inversely related to the distance between the NS 5-branes 
as well as directly related to 
the IIA string coupling constant, $g_s$.
Since the D2-brane tension is much smaller than the D4 brane 
tension, the gauge symmetry on the D4 branes appears 
as a global symmetry of the 1+1 dimensional theory on the 
intersection. 
We extract quantum information about these 
two dimensional gauge theories by making 
the IIA string coupling constant large; that is, 
we embed the ten dimensional IIA brane construction into 
eleven dimensional 
M-theory. 
We find that the metric on the 
Coulomb branch of the $U(N_c)$ gauge theory 
receives quantum corrections, deforming it 
to the so called wormhole metric of 
\callan. This result was found by field theory techniques
in \natidia. We also find that there is another theory,
$\prod_i U(n_i)$ with bi-fundamental matter fields, 
having an ALE-type metric with $\theta = 0$ which 
flows to the same wormhole metric in the IR.  This duality 
is reminiscent of Seiberg's ``non-Abelian Coulomb phase'' 
in four dimensions \neqone\ and 
mirror symmetry in two dimensions \mp.
In this respect, this duality is different from the mirror symmetry 
found in three dimensional gauge theories which 
relates strong to weak coupling.

In section two, we present the type IIA brane construction 
of $N = (4,4)$ two dimensional gauge theories. In section three,
we study quantum effects on the world volume theory of the 
D2-brane by considering 
the bending of the branes in IIA and by making the eleventh 
dimension very large. 
The torsion of the Coulomb branch moduli space is 
seen as coming from the field strength of the 
self-dual 2-form field of the $(0,2)$ theory 
on the world volume of the 5-brane. In section four, we review 
mirror symmetry in three dimensions and then consider compactification 
of this symmetry to two dimensions. In the process we review Buscher's 
duality for the wormhole metric and the Taub-NUT metric 
and relate it to transformations in the brane construction. In 
section 6, we examine the monopole moduli space of five dimensional 
gauge theories
and find two dual interpretations. In section 
7, we consider two dimensional gauge 
theories with an adjoint hypermultiplet in 
addition to fundamental hypermultiplets, and 
we relate the two dimensional mirror symmetry to the 
duality between the $(1,1)$ and $(0,2)$ string theories. 
In section 8, we propose that there is a Seiberg-type duality 
in two dimensions that relates different Higgs branchs to 
the same conformal field theory. We end by speculating about 
a quantum Higgs branch.

\newsec{Field Theory on the D2 brane}
\subsec{The Brane Construction}
\subseclab{\construct}

The configurations we will study involve three
kinds of branes in type IIA string theory: 
a Neveu-Schwarz (NS) fivebrane,
Dirichlet (D) fourbrane and Dirichlet 
twobrane.
Specifically, the branes are:

\item{(1)} NS fivebrane with worldvolume 
$(x^0, x^1, x^2, x^3, x^4, x^5)$ lives at 
a point in the $(x^6, x^7, x^8, x^9)$ directions.
The NS fivebrane preserves supercharges of the 
form\foot{$Q_L$, $Q_R$ are the left and right moving 
supercharges of type IIA string 
theory in ten dimensions. They are (anti-) chiral: 
$\epsilon_R=-\Gamma^0\cdots\Gamma^9\epsilon_R$,
$\epsilon_L=\Gamma^0\cdots\Gamma^9\epsilon_L$.}
$\epsilon_LQ_L+\epsilon_RQ_R$, with 
\eqn\nsfive{
\eqalign{\epsilon_L=&\Gamma^0\cdots\Gamma^5\epsilon_L\cr
\epsilon_R=&\Gamma^0\cdots\Gamma^5\epsilon_R.\cr
}}

\item{(2)} D fourbrane with worldvolume
$(x^0, x^1, x^7, x^8, x^9)$ lives at a point
in the $(x^2, x^3, x^4, x^5, x^6)$ directions.
The D fourbrane preserves supercharges satisfying
\eqn\dfour{
\epsilon_L=\Gamma^0\Gamma^1\Gamma^7\Gamma^8
\Gamma^9\epsilon_R.
}

\item{(3)}
D twobrane with worldvolume $(x^0, x^1, x^6)$
lives at a point
in the $(x^2, x^3, x^4, x^5, x^7,x^8,x^9)$ directions.
The D twobrane preserves supercharges satisfying 
\eqn\dtwo{
\epsilon_L=\Gamma^0\Gamma^1\Gamma^6\epsilon_R.
}

It is easy to check that there are eight real supercharges
satisfying equations \nsfive-\dtwo,
${1\over 4}$ of the original supersymmetry of 
type IIA string theory. Each relation \nsfive-\dtwo\ by itself would break
${1\over 2}$ of the supersymmetry. Equations \nsfive\ and 
\dfour\ are independent and together break to ${1\over 4}$. 
Equation \dtwo\ is not independent of \nsfive\ and 
\dfour\ and hence breaks no more of the supersymmetry.
Altogether the branes preserve
${1\over 4}$ of the supercharges. 
If one T-dualizes along the $x^2$ direction, one recovers the 
IIB construction 
of \hw.

\subsec{The Fields}
\subseclab{\fields}

The global R-symmetries of the $N = (4,4)$ supersymmetric theory on the 
D2-brane arise from the Lorentz group of the 
ten dimensional space-time. We have a 
$Spin(4)$ symmetry in the directions $x^2,x^3,x^4,x^5$ and an 
$SU(2)_R$ symmetry of the 
coordinates $x^7,x^8,x^9$. 
There is also the $SO(1,1)$ Lorentz symmetry of the coordinates
$x^0,x^1$ which is space-time for the two dimensional theories 
which we will be interested. 
In all we have $SO(1,1)\times Spin(4) \times SU(2)_R$.
These are the correct symmetries for the $N = (4,4)$ 
two dimensional theory \natidia.
The ten dimensional 
$\bf {16^{+}}$ and $\bf {16^{-}}$ supercharges 
of the IIA string theory leave unbroken 
$\bf{(1,2,2)^{+} \oplus (1,2,2)^{-}}$ 
under the  $SO(1,1)\times Spin(4) \times SU(2)_R$.

Hypermultiplets in the fundamental representation of the 
gauge group arise from strings stretched between the 
D2-branes and the D4 branes. Scalars in these hypermultiplets 
transform as $\bf {(1,1,2)} \oplus \bf {(1,1,2)}$ 
under $SO(1,1)\times Spin(4) \times SU(2)_R$. 
We can understand this as follows: in the brane construction
Higgsing corresponds to breaking a D2-brane between 
two D4 branes and moving it in the $x^7,x^8,x^9$ direction. 
We also must include the component $A_6$ from the gauge field, 
which is the only surviving part of the gauge field permitted by these 
boundary conditions. Once we embed 
the IIA theory in M-theory, the 
$x^{10}$ replaces $A_6$ as the forth scalar. 
We see the scalars of the Higgs branch 
naturally fall into representations  
$\bf {3} \oplus \bf {1}$ under the $SU(2)_R$ corresponding to the three 
complex structures of the hyper-Kahler Higgs branch.

In terms of $N = (2,2)$ superfields,
the Coulomb branch consists of 
twisted (vector) multiplets,
$\Lambda $,
and normal chiral multiplets,
$\Phi $.
The 
$U(1)$ gauge field on the D2-brane 
lives in the twisted multiplet, and motion in 
the directions 
$x^2,x^3,x^4,x^5$
corresponds to the scalar fields 
components of both
$\Phi$ and $\Lambda$. 
These scalars 
are therefore charged under the $\bf{(1,4,1)}$
of the $SO(1,1)\times Spin(4) \times SU(2)_R$.
The Coulomb branch, parameterized 
by the scalars, is characterized by 
a Kahler potential which satisfies 
a four-dimensional Laplacian 
\eqn\Laplace{\d_{\Lambda}\d_{\bar\Lambda}K + \d_{\Phi}\d_{\bar\Phi}K=0.}
This determines the metric 
\eqn\metric{ds^2=\d_{\Phi}\d_{\bar\Phi}K d\Phi d\bar\Phi -
\d_{\Lambda}\d_{\bar\Lambda}K d\Lambda d\bar\Lambda}
and the antisymmetric tensor field 
\eqn\fftor{B={1 \over 4} (\d_{\Phi}\d_{\bar\Lambda}K d\Phi\wedge
d\bar\Lambda+\d_{\Lambda}\d_{\bar\Phi}K d\bar\Phi\wedge d\Lambda).}
The torsion is determined from $H = dB$. 

It has been argued that the Coulomb branch and the 
Higgs branch flow to different and distinct theories 
in the infra-red \ed. The basis for this claim is 
that the superconformal theory which these 
gauge theories flow is known to have an 
$SU(2) \times SU(2)$ global symmetry. Since
scalars must be singlets under this symmetry,
the Higgs branch must flow to a theory with global symmetry 
coming from the $Spin(4)$ since it has scalars charged as 
${\bf (1,1,3)\oplus (1,1,1)}$ under  
$SO(1,1)\times Spin(4) \times SU(2)_R$. Likewise, 
the Coulomb branch must flow to a theory with global symmetry 
coming from the $SU(2)_R$ since it has scalars charged as 
${\bf (1,4,1)}$ under  
$SO(1,1)\times Spin(4) \times SU(2)_R$.
One must keep in mind that there is no moduli space in 
two dimensions and we will always be working 
in the Born-Oppenheimer approximation.

\subsec{The Parameters}
\subseclab{\parameters}

Giving a mass to a 
hypermultiplet in the brane picture corresponds 
to moving a D4 brane away from a D2-brane in the 
$x^2,x^3,x^4,x^5$ direction. Therefore, masses transform in 
the $\bf{(1,4,1)}$ of the $SO(1,1)\times Spin(4) \times SU(2)_R$
symmetries. 
The Fayet-Illiopoulos parameters, $\vec \zeta$, 
correspond to motion of 
the NS 5-branes in the $x^7,x^8,x^9$ direction, and therefore 
they transform in the 
$\bf{(1,1,3)}$ of the $SO(1,1)\times Spin(4) \times SU(2)_R$
symmetries. They are background hypermultiplets. 
The $\theta$ angle corresponds to the 5-th scalar of the $(0,2)$ 
tensor multiplet on the worldvolume theory of the NS 5-brane.
As will be discussed below, going to M-theory makes the 
$\theta$ angle manifest.
The FI parameters can be viewed as the Kahler deformations 
of the metric of the associated 
hyper-Kahler non-linear sigma model \phases.
Adding the theta angle gives a ``quaternionic Kahler form''
$\vec \zeta + i\theta$, analogous to the complexified 
Kahler form used in $N = (2,2)$ theories. 
The $\theta$ parameter is to be associated with deformations of 
the antisymmetric tensor 2-form field of the 
non-linear sigma model. 
The distance between the NS 5-branes in the $x^6$ direction is 
inversely proportional to the gauge coupling constant
\eqn\coupling{{1\over g_2^2} = {x^6\over M_sg_s}}
where $M_s$ is the string scale.

\newsec{Quantum Effects on the Brane}
\subsec{Hanany-Witten Transition}
\subseclab{\transition}

We will consider a $U(N_c)$ gauge theory with $N_f$ 
fundamental hypermultiplets.
The corresponding brane configuration has $N_c$ D2-branes suspended 
between two NS 5-branes separated in the $x^6$ direction 
with $N_f$ D4-branes intersecting the 
D2-branes at points in $x^6$.
One can move the D4 branes in the $x^6$ direction 
outside of the NS 5-branes. In doing so, $N_f$
D2-branes are created. Strings between the D2-branes 
inside the NS 5-branes and the newly created D2-branes outside 
give rise to the hypermultiplets.
Situations with different $x^6$ positions for the D4 branes 
have been found to be equivalent
in higher dimensions \hw. 
Since the two dimensional case is 
related to higher dimensional cases 
by T-duality,
we will assume different $x^6$ positions are equivalent here (see Figure 1).
\ifig\Fone{The Hanany-Witten transition. The horizontal direction 
is $x^6$. Both diagrams are equivalent from   
field theory.}
{\epsfxsize4.0in\epsfbox{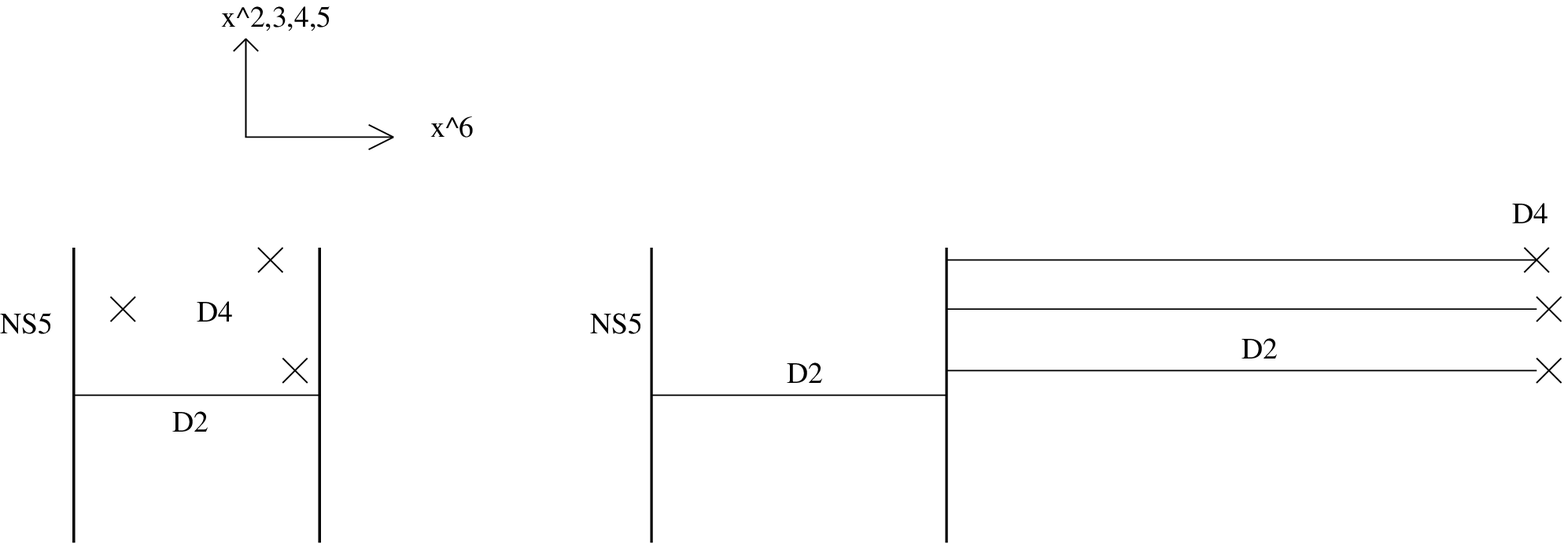}}

Once the $N_f$ D4 branes are outside of the NS 5-branes, we 
can consider what happens when the $N_f$ D2-branes start to pull 
on the 5-branes. The tension of the D2-branes 
inside the NS 5-branes is negligible in comparison
to the D2-branes on the outside since
they are much shorter, and therefore we will not 
consider their effect on the deformation of the 5-brane.
Since the tension of the NS 5-brane goes as ${1\over g_s^2}$, 
bending of the branes is quantum mechanical in string theory.
This can be seen by taking $g_s \rightarrow 0$. In this limit,
the tension of the branes goes to infinity and there can be 
no bending. This is the classical limit for the world 
volume theory on the D-branes 
(which are non-perturbative objects in string theory).
Quantum information about the field theory can therefore be 
gained from considering branes that bend (see Figure 2). 
\ifig\Ftwo{A D2-brane pulls on the D5-brane, bending it as shown. 
The D2-brane doing the pulling becomes tensionless since there is 
nothing to balance it's tendency to collapse.}
{\epsfxsize3.0in\epsfbox{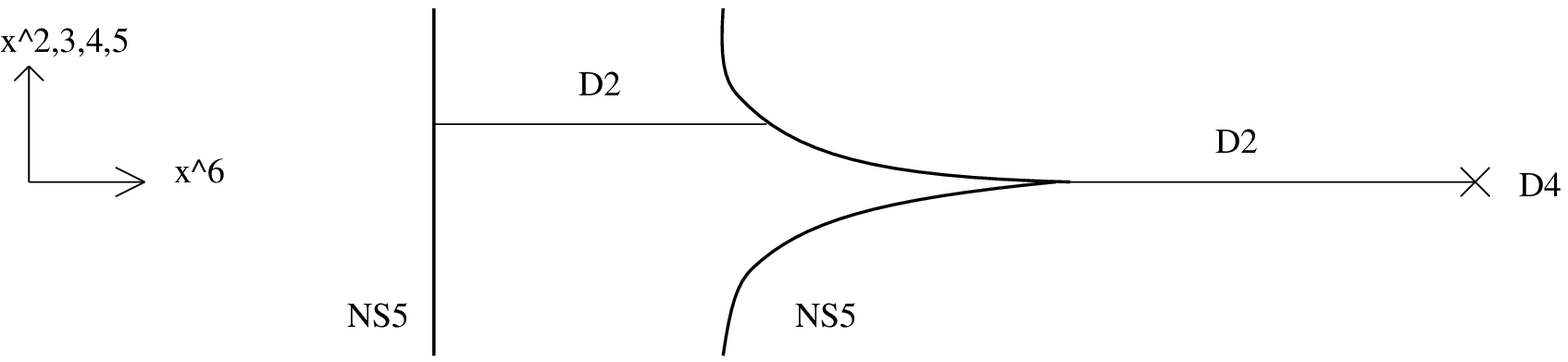}}
The physical situation of a D2-brane 
pulling on a NS 5-brane satisfies the 
four dimensional Laplacian, $\nabla^2 x^6 = \delta^{(4)}$, where the 
four dimensions are the ones on the 5-brane
transverse to the 2-brane. 
A solution of the
Laplacian is $x^6 = a/ r^2 + c$ where 
$r^2 = (x^2)^2 + (x^3)^2  + (x^4)^2  + (x^5)^2$. 
This solution minimizes the 
surface area of the 5-branes. 
We have
for gauge group $U(1)$, 
\eqn\meta{ds^2 = \left({1\over g_{2}^2}
+{N_f\over r^2}\right)(dr^2 + r^2d\Omega_3^2)}
where $r$ is now a coordinate on the moduli space.
\foot{The coordinates on the moduli space and 
the coordinates in IIA space-time are related by $M_s^2$, the string scale.}
In field theory the modification to the flat metric 
comes from a one-loop calculation.
There is also the antisymmetric tensor field
\eqn\Tora{B=-{1 \over 4} N_f \sin^2{\theta\over 2}d\phi\wedge d\chi}
where $0\leq\theta <\pi,\,0\leq\phi <2\pi,\,0\leq\chi <4\pi$
are angular coordinates on the unit three
sphere $S^3\subset R^4$
\eqn\angcoord{
\Phi= e^{i(\chi-\phi)/2}\cos{\theta\over 2},\qquad
\Lambda= e^{i(\chi+\phi)/2}\sin{\theta\over 2}.\qquad}
$H$ is non-zero when 
the $B$-field is not well defined on $S^3$ giving 
the moduli space torsion.
We will see below that \Tora\ can 
naturally be associated with the $B_{\mu\nu}$ field of the 
worldvolume theory on 
the NS 5-branes of IIA. 
What we 
see is that as the 2-brane moves on the 5-brane it encounters
singularities at the points where there is another 2-brane pulling 
on the 5-brane from the other side. This is the analogue 
of the $log$ singularity found in the 4d case in \bending. 
The differences here are that the $1/r^2$ singularity is 
not corrected by instantons and that the singularity is infinitely far away
from any point on the Coulomb branch. This metric \meta\ is identical 
to the wormhole solution of the NS 5-brane found in \callan.
From the perspective of \callan, $N_f$ is the amount of 
charge coupling to the field strength $H = dB$, the torsion.

What is intriguing about the situation here is that the 
D4 branes can actually be at finite distance from the 
NS 5-branes. In this configuration 
it is possible to ``go down'' the throat of the 
wormhole and see what is at the bottom. We find that 
there is a tensionless D2-brane where the Coulomb 
branch meets the Higgs branch. This is consistent 
with there being an 
non-trivial fixed point at the origin of the 
Coulomb and Higgs branches; when the world volume of 
the brane probe comes into contact with a brane of zero 
tension, there is often a non-trivial fixed point \oferami.
Presumably, this is related to the new massless states 
that arise when a brane becomes tensionless and 
the fact that the theory becomes scale invariant as $g_{cl} \rightarrow \infty$.

It was shown \callan\ that the non-linear sigma model 
having the 5-brane as a target space is a 
level $N_f$ supersymmetric $SU(2)$ WZW model 
with a Louiville field associated with 
the dilaton.
Near the singularity, 
the metric \meta\ can be rewritten as 
\eqn\metrici{ds^2=
{N_f\over r^2}\left(dr^2+r^2d\Omega_3^2 \right).}  
We can introduce a new radial coordinate
$\eta{=}\sqrt{N_f} {\rm log}(r/\sqrt{N_f})$ to give:
\eqn\metricii{ds^2=
\left(d\eta^2+N_fd\Omega_3^2 \right),} 
The field $\eta $ blows up as we proceed down the 
throat of the wormhole. Consequently, the Louiville field
and the dilaton also blows up. This is 
therefore a singular conformal field theory. 
The topology here is ${\bf R^{+} \times S^3}$.
In the brane picture, as the D2-brane extends down the 
wormhole, the theory essentially becomes three dimensional. 
This is consistent with the idea \edINamst\ that the 
extra dimension of the D2-brane, $x^6$, 
can be associated with a Louiville field.
Since we find new massless states at the bottom of the throat, we 
conclude that there should perhaps
be another description that replaces the CFT of \callan, 
in the limit that the dilaton blows up.

\subsec{Torsion}
\subseclab{\torsion}

We interpret the torsion of the moduli space of the Coulomb branch in the 
brane picture in the following way. When $N_f$ D2-branes end on 
an NS 5-brane of type IIA string theory, the D2-branes look like strings 
in a 5+1 dimensional theory. Strings in six dimensions 
couple to the self-dual 2-form field, $B_{\mu\nu}$. The 
charge that the strings carry is therefore 
$N_f = \int_{S^3} H$ where $H^{(3)} = dB$.
In two dimensional sigma models, torsion 
comes from a $B_{\mu\nu}$ field and is equal to 
$H^{(3)}$ \rocek. Each fundamental 
hypermultiplet, contributes an integer amount 
to the torsion of the Coulomb branch.
We therefore identify \fftor\ with the 
2-form field of the six-dimensional 
$(0,2)$ theory. 

A D2-brane ending on a D4-brane 
also looks like a string, but in 4+1 dimensions. 
The string couples electrically to the $B_{\mu\nu}$ 
field, which is the dual of the photon on the worldvolume theory
on D4 brane. The magnetic charge carried by each 
string is therefore $N_f = \int_{S^2} F^{(2)}$ 
where $F = dA$. Each D2-brane looks like a monopole in 
the D4-brane. Monopole moduli space is given by a Taub-NUT metric
which is torsion free. 
This is consistent with there being 
no torsion on the Higgs branch since motion of a D2-brane
between two D4-branes corresponds to Higgs branch moduli.
We will see below that in M-theory the 
metric on the Higgs branch 
is better thought of as being T-dual to the Taub-NUT.

This analysis is also consistent with there being no torsion 
for the Coulomb branch in three dimensions. The three dimensional 
Coulomb branch is described 
by a D3-brane ending on an NS 5-brane in IIB string theory.
The D3-brane looks like a 2-brane in 5+1 dimensions which 
couples electrically to a 3-form field $C^{(3)} = *A^{(1)}$, the 
dual of the photon in six dimensions. The D3 brane is 
therefore a monopole on the NS 5-brane.
There is no coupling to a $B_{\mu\nu}$ field, $H=0$, 
and the Coulomb moduli space 
has no torsion. Likewise, 
a D4-brane ending on an NS 5-brane of IIA describes 
Coulomb branch of the four dimensional theory.
There is no torsion here either 
since the D4-brane, being a vortex solution in the 
5+1 theory, does not couple to the 2-form field. Again 
this is consistent with there being no torsion on
the Coulomb branch moduli space of theories in four
space-time dimensions.

\subsec{The view from M-theory}
\subseclab{\m}

In this section, we look at the brane configuration in M-theory 
and attempt to understand the infra-red dynamics 
of the associated gauge theory.
Let's start in IIA where the compact direction $x^{10}$ is very small. 
As explained in section \parameters, motion of the NS 5-branes 
in the $x^7,x^8,x^9$ direction corresponds to 
an FI term $\vec \zeta$. Once the direction $x^{10}$ becomes of finite radius, 
there is a new direction in which the 5-brane can move. This is 
the theta angle, $\theta$. Notice that a non-zero 
theta angle breaks supersymmetry if all the hypermultiplets have 
zero vacuum expectation value $<Q> = 0$. This is consistent 
with the interpretation of the theta angle as an electric field 
since it creates a non-zero vacuum energy.
\foot{This was independently 
noted in \amihori.} 
As was discussed in \somecomments, the point where $<Q> = <\Phi > = 
\vec \zeta = \theta = 0$
is the point where the Higgs branch meets the Coulomb branch. Therefore,
the wavefunction on the Higgs branch can move onto the Coulomb branch. 
In \somecomments, this was interpreted as evidence for the 
conformal field theory on the Higgs branch becoming singular. 

What is interesting, now that the eleveth dimension 
has become big, is that we can see that the 
$$SO(1,1)\times Spin(4) \times SU(2)_R$$ global symmetries have
become enhanced to $$SO(1,1)\times Spin(4)_H \times Spin(4)_C.$$
In \ed\ it was argued that the theory on the Coulomb branch decouples 
from the theory on the Higgs branch and flows to a superconformal 
fixed point with an $SU(2) \times SU(2)$ symmetry. Since the 
scalars of the Coulomb branch transform under the $Spin(4)_H$, this is not a 
candidate. It was conjectured in \ed\ that the $SU(2)_R$ gets 
enhanced to $Spin(4)_C$. Here we see a realization of that idea; the 
opening up of the eleventh dimension makes the Lorentz group larger
which appears in the brane construction as an enhanced R-symmetry. 
What's more, we see that there is a natural symmetry 
between the $Spin(4)_H$ and the $Spin(4)_C$ 
exchanging the Coulomb and Higgs branches, and the mass parameters of 
the fundamental fields with the FI parameters and theta angle.
This is much like the relationship between the R-symmetries in 
three dimensions between the $SU(2)_H$ of the Coulomb branch 
and the $SU(2)_C$ of the Higgs branch. Notice, however, there
is an important difference. There is no S-duality in M-theory, as
there was in IIB theory, that enables us to transform the 
Higgs branch into the Coulomb branch as was done in \hw. 
There is however 
another IIA theory with a 
``mirror'' Higgs branch and Coulomb branch that 
becomes equivalent to this theory in eleven-dimensional M-theory. 

The two IIA brane configurations that flow to the 
same configuration in M-theory are in fact the 
same ``mirror'' pairs that were discussed in three 
dimensions in \kennati.
For example, a $U(1)$ gauge theory with $N_f$ 
flavors is ``mirror'' to a $U(1)^{N_f -1}$ gauge theory
with matter charged in the $(1,0,0,0,0,...,0)$, 
$(1,-1,0,0,0,...,0)$,
$(-1,1,0,0,0,...,0)$, $(0,1,-1,0,0,...,0)$, 
$(0,-1,1,0,0,...,0)$, ...., $(0,0,0,0,0,...,0,1)$.
The Higgs branch of this mirror theory 
is classically an $A_{N_f-1}$ ALE space.
However, unlike usual orbifold theories
here we have $\theta = 0$ since we consider no 
separation between the 5-branes in the 
$x^{10}$ direction.
This is a singular limit in the conformal 
field theory describing the ALE space \refs{\asp, \somecomments}.
As discussed above, the $U(1)$ theory with 
$N_f$ flavors is constructed by suspending 
a D2-brane between two NS 5-branes with $N_f$ 
D4-branes intersecting the D2-brane at points in the $x^6$ direction.
The mirror theory is constructed
by suspending a D2-brane between two D4-branes with 
$N_f$ NS 5-branes intersecting the D2-brane at points in the 
$x^6$ direction. Upon going to M-theory, 
the D4-branes become 5-branes, the NS 5-branes remain 5-branes, 
and the D2-branes remain membranes. Therefore, the two constructions 
which were different in IIA string theory are equivalent in M-theory. 

\newsec{Compactification of the theory from three dimensions to 
two dimensions.}
\subsec{Review of 3d mirror symmetry}
\subseclab{\review}

Here we review the mirror symmetry of \kennati, 
and in the next section explore it's reduction to 
two dimensions. 
The three dimensional construction 
of the branes was carried out in \hw. It is equivalent to our construction 
in section \fields\ once we T-dualize upon $x^2$. 
We start in IIB string theory 
with $N_c$ D3 branes in directions $(x^0,x^1,x^2,x^6)$ stretched 
between NS 5-branes in directions $(x^0,x^1,x^2,x^3,x^4,x^5)$. 
Motion of the D3 branes in the 
$x^3,x^4,x^5$ direction constitutes motion on the Coulomb branch. 
The Higgs branch is provided by $N_f$ D5 branes 
in $(x^0,x^1,x^2,x^7,x^8,x^9)$ which allows the D3 brane to 
break on the D5 branes and move in the $x^7,x^8,x^9$ direction. 
The fourth scalar of the quaternionic Higgs branch is provided by the 
$A_6$ component of the gauge field. Let's consider, for definiteness,
a $U(1)$ theory with $N_f = 1$ although the following statements
will be fairly general. We have one D3 brane suspended between 
two NS 5-branes and one D5 brane intersecting the D3-brane at a point in $x^6$. 
We consider performing the Hanany-Witten transition and putting the D5 brane 
outside of the NS 5-branes, creating a new D3 brane stretched between the 
NS brane and the D5-brane. 
The D3 branes pull on the NS 5-branes creating a $1/r$ singularity. As before, 
we interpret this as meaning that the Coulomb branch has an $1/r$ singularity.
If we dualize the photon we get a compact scalar. 
In the IIB string theory, dualizing the photon 
corresponds to performing an S-duality which
turns the NS 5-branes into D5-branes and takes strong to weak coupling. 
The dual photon is the $A_6$ component 
of the D3 brane which is not projected out by the 
D5 brane boundary conditions, 
while the fields $A_1$ and $A_2$ are projected out. In the 
dual variables the Coulomb branch is $\bf {R^3 \times S^1}$, the $S^1$ 
coming form the compact scalar $A_6$. 
Quantum mechanically, the Coulomb branch is modified to a Taub-NUT metric. 
\eqn\TNmet{ds^2=g_3^{2}(\vec x)(d\sigma+\vec\omega\cdot d\vec
x)^2+g_3^{-2}(\vec x)d\vec x\cdot d\vec x,} with
\eqn\gis{g_3^{-2}(\vec x)=g_{cl}^{-2}+\sum _{i=1}^{N_f}{1\over |\vec x-\vec
m_i|},\qquad \vec{\grad {}}(g_3^{-2})=\vec{\grad {}}\times \vec \omega.} 
where $\vec x = (x^3,x^4,x^5)$ and $\sigma$ is the compact 
direction $A_6$.
For $N_f = 1$ there is a removable singularity in the metric.
In the brane configuration, we now have D3-branes pulling on 
a D5-brane. The singularity due to the 
bending is just \gis\ where $g_{cl}$ is the separation 
between the D5 branes in $x^6$ far from the singularity. 
As we approach the singularity, 
the effective coupling of the three dimensional gauge theory goes to zero 
since $x_6$ goes to infinity. 
Since the radius of the $S^1$ fiber of the Taub-NUT metric 
\TNmet\ is proportional 
to $g_3$, the fiber shrinks to zero radius at $r=0$.
In this limit, we can neglect the 
constant $c$, and the metric goes over to the ALE metric (see Figure 3).
\ifig\Fthree{We see how the bending of the D5-branes creates a Taub-NUT metric.
 $S^2$ indicates that there is a 2-cycle whose radius depends 
on the distance between the NS 5-branes (the FI terms).}
{\epsfxsize3.0in\epsfbox{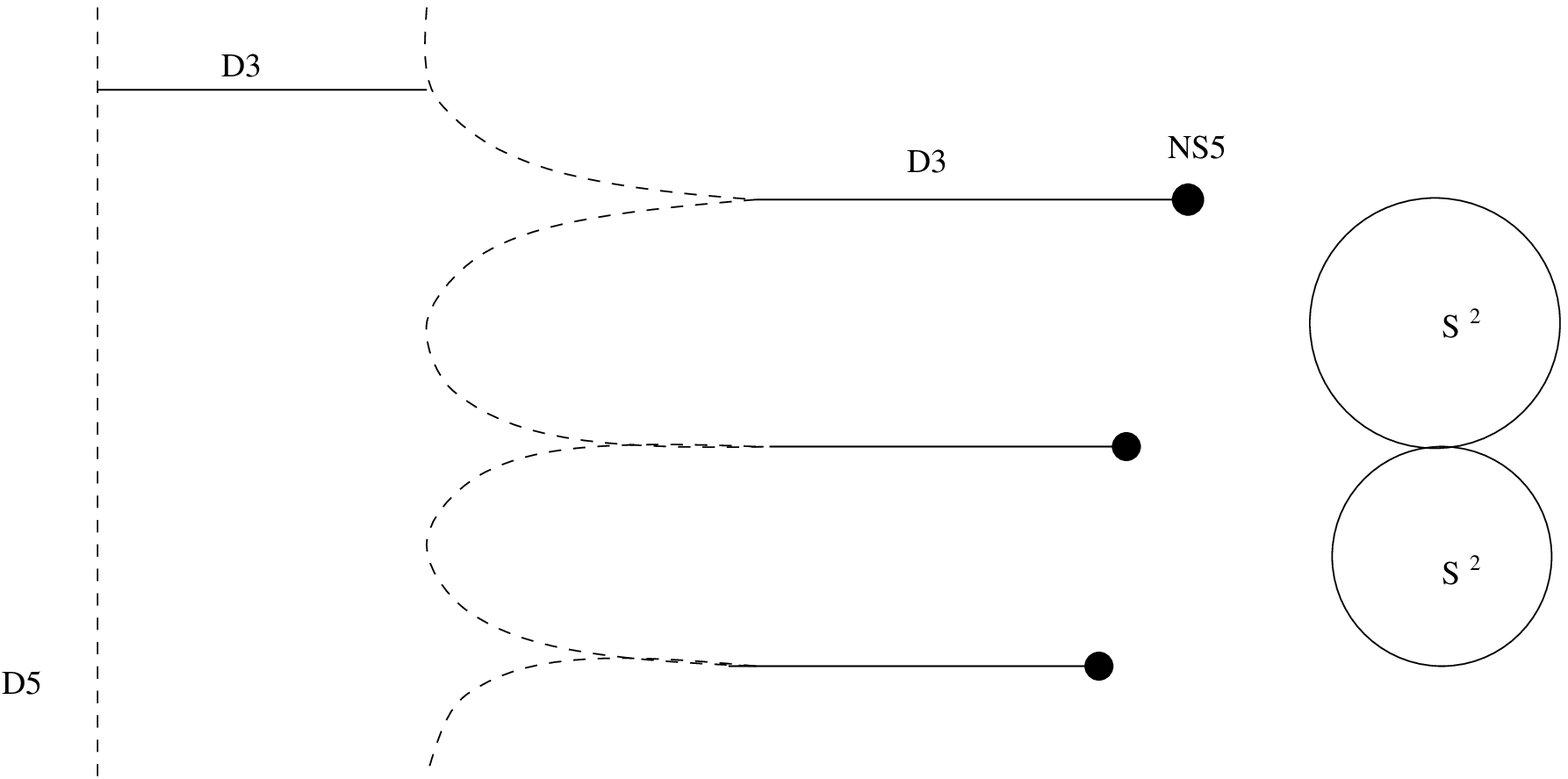}}

We saw above, that the singularity at the origin of the Coulomb
branch has a description in terms of dual (magnetic) coordinates 
as a Taub-NUT metric.
As $g_{cl}$ goes to infinity, the singularity is also 
described by
an ALE singularity of a classical mirror Higgs branch, which from 
the perspective of the branes is the same as the dual coordinates 
of the Coulomb branch. This is seen as 
evidence \kennati\ for a non-trivial fixed point at the origin of the Higgs 
and Coulomb branches for $N_f > 1$ for a $U(1)$ gauge theory. 
At $r = 0$, it appears that there 
is a tensionless D3 brane for reasons similar to those 
argued in section \transition.
This would be consistent with there being a non-trivial fixed point 
at the origin of the Higgs and Coulomb branches \oferami.

\subsec{Buscher duality.}
\subseclab{\buscherDUAL}

It is well known that if one T-dualizes along the $S^1$ of the Taub-NUT 
space one gets a wormhole metric 
with one of it's transverse directions compactified (also 
called an H-monopole)
\refs{\ghm, \banks}. Let's review this duality.
In the metric \TNmet, the direction $\sigma$ is compact. 
This is an isometry; we can dualize upon this. Rewriting \TNmet
\eqn\TNbusch{ds^2 = 
g_3^{-2}(\vec x)d\sigma d\sigma + g_3^{2}(\vec x)\omega_id\sigma 
dx^i + g_3^{2}(\vec x)\omega_i dx^i d\sigma + 
(g_3^{2}(\vec x)\omega_i \omega_j + g_3^{-2}(\vec x)\delta_{ij})
dx^i dx^j}
Gauging the isometry and then integrating out the gauge fields as 
described in \rv\ and \buscher, we find that 
\eqn\WHbusch{\eqalign{
ds^2 = & g_3^{2}(\vec x)(d\sigma d\sigma + \delta_{ij} dx^i dx^j) \cr
b_{\sigma i} = & \omega_i \cr
\Phi = & log(g_3^{-2}(\vec x)) \cr  
= & log ( g_{cl}^{-2}+\sum _{i=1}^{N_f}{1\over |\vec x-\vec
m_i|}  ) \cr}}
The $log$ appears because under T-duality the ratio ${e^{2\Phi}\over R}$
must be preserved.
We see that the cross-terms in \TNbusch\ have produced the 
antisymmetric tensor field 
of \WHbusch. Moreover, we have produced the logarithmic dilaton 
and a metric that has the same form as the wormhole solution described in 
\meta \callan. 
However, there is a difference between this metric \WHbusch\ and the 
one in \meta; the $g_3^{-2}(\vec x)$ given in 
\TNmet\ goes as $1/|\vec x|$ whereas the wormhole metric of 
\callan\ goes as $1/r^2$. The difference is clearly due to the 
fact that \WHbusch\ the direction $\sigma$, 
which is transverse to the 5-brane, 
is compact, and therefore the appropriate Laplacian is three dimensional 
rather than four dimensional. 
Moreover, we are neglecting states that propagate in the compact 
direction. When we decompactify the $S^1$,  
the $g_3^{-2}(\vec x)$ of \WHbusch\ becomes $1/|\vec x|^2 + |\sigma|^2$, 
we must include states that propagate in the compact direction,
and, as noted in \natidia, the wormhole 
metric no longer has an isometry and therefore
we do not know how to dualize. Momentum modes that probe the wormhole 
metric 
are dual to strings that wind around the $S^1$ of the Taub-NUT. 
In \ghm, it was found that the winding modes should see a more ``throat-like''
behavior of the Taub-NUT. 

\subsec{Dimensional reduction from 3 to 2}

In \natidia, it was conjectured that 
the wormhole metric on the Coulomb branch 
has another interpretation in terms of some 
unspecified dual coordinates, where the 
metric is ALE. We will here attempt to understand this 
speculation in terms of brane constructions.
Consider compactifying the direction $x^2$ of the 
IIA set up in section \fields\ and 
wrapping the NS 5-brane on an $S^1$ of small 
radius $R_2$. The bending of the branes 
gives us the metric \WHbusch\ with a $1/r$ singularity.
By T-duality, we can 
turn this into a large circle of radius $R_B = 1/M_s^2R_2$
which also takes the D2-brane to a D3 brane, and a IIA 
5-brane with a $B$ field to a IIB 5-brane without a B-field.
This is the set-up of \review.
Although from the point of view of the NS 5-brane the D3 brane is 
a monopole and the moduli space of monopoles should 
have a Taub-NUT metric, in field theory the Taub-NUT is only
visible once we dualize the compact scalar. In the brane theory
that corresponds to an S-duality of IIB.

In the limit that $R_2$ becomes big, the description 
of the metric in the IIA theory is better described by 
\metrici\ which has a $1/r^2$ singularity. T-dualizing this 
to IIB, we have a D3 brane wrapped on a small circle 
with radius $R_B$. This is a 3d theory 
on a $R^{1,1} \times S^1$ base space where the scalar
parameterizing the Wilson loop is big. 
If we now dualize the photon, which corresponds to 
performing S-duality on the IIB configuration, this 
takes us to a configuration of D5 branes and D3 branes wrapped 
on $x^2$. From the perspective of the D5 brane, the D3 branes are monopoles.
Therefore, the metric on the moduli space of D3 branes is given by 
a Taub-NUT metric \TNmet\ which in the limit that $g_{cl}$ 
goes to infinity becomes the ALE metric. 
This appears to be a realization of the speculation in 
\natidia\ that there are some coordinates in which 
the wormhole metric is ALE.
However, 
once we T-dualize on $x^2$, taking the S-dual IIB configuration 
to IIA, 
we find a configuration of D2-branes suspended between 
D4 branes. 
The metric here is also ALE with 
$\theta = 0$. This sigma model has no well defined conformal 
field theory.
As we open up the eleventh dimension, the fourth scalar of the 
Higgs branch no longer comes from $A_6$ but rather from $x^{10}$.
The metric is now better thought of as the metric \WHbusch\ rather 
than ALE.
As the radius of the eleventh dimension becomes large, 
the D4 brane decompactifies into a 5-brane.
The fact that we can give non-zero values to the fourth scalar, $x^{10}$, 
and parameter, $\theta$,  
implies that there should be four parameters, $\vec \zeta + i\theta$, 
to 
tune such that there is a singularity, unlike \TNmet\ which 
only has three parameters to tune.
It is clear that this must be the case, since 
as the 11-th dimensions opens up, we see that the Laplacian of the 
D2-brane pulling on the D4-brane, goes from being 
a three dimensional Laplacian to being 
the four-dimensional Laplacian of a D2-brane pulling on a 5-brane. 
The metric on the Higgs branch is therefore
\meta. We said before that deformations 
of the antisymmetric 2-form of the non-linear sigma model 
are to be associated with $\theta$, the distance between the 5-branes in 
the eleventh dimension. Allowing 
the eleventh dimension to open up, allows for a new deformation 
to the corresponding non-linear sigma model, the antisymmetric 
tensor field. $H$ is non-zero at the point where $B$ is not well defined. 
This point where $H$ is non-zero 
occurs in the branes when $\theta$ becomes non-compact. 
This is consistent with the appearance of torsion in the metric \meta.

Since the Higgs branch of the mirror 
theory is ALE, it is hyperKahler and such sigma models do not receive 
quantum corrections. 
On the other hand, since we 
are considering the point where there is no well defined CFT
, $\theta = 0$, 
it is not clear that the non-renormalization theorems apply. 
Moreover, in \somecomments, it was conjectured that 
the ALE orbifold theory with $\theta = 0$ flows to the 
CFT of \callan, discussed in section \transition.
Here we see a realization of that idea and a resolution of a puzzle:
In \somecomments, it was not clear how the ALE metric would develop 
torsion. Here we see that torsion comes about when the gauge field of the 
4+1 theory becomes the self-dual 2-form of the $(0,2)$ 5+1 dimensional 
theory at strong coupling. 
Since the D2 branes in the 5-brane 
look like strings charged under $H$, the moduli 
space develops non-zero torsion.
Branes allow us to see such novel non-perturbative 
effects explicitly.

\newsec{Monopoles in 4+1 SYM}
If the theory of 
D2-branes suspended between D4 branes is, from the D4 brane perspective, 
a monopole moduli space, what happens when we go to M-theory? What 
happens to the monopoles?
The answer is that the 4+1 Super Yang-Mills theory is not well defined; 
it flows to a free theory in the IR. In order to 
have a well defined theory, we must embed the 4+1 SYM into the (0,2) 5+1 
theory that was described in \newtheories \sbr. If we want quantum information 
about the 4+1 SYM, we must consider the 5+1 (0,2) theory. Magnetically 
charged objects in the 4+1 SYM theory are strings. When we go to the 
(0,2) theory, the strings remain strings. However, from the 
point of view of the 5+1 theory, the strings are co-dimension four objects.
The string moduli space of the 5+1 (0,2) theory 
is the moduli space of monopoles of the 4+1 SYM. Therefore,
the metric on the moduli space of monopoles should be something 
that interpolates between $1/r$ and $1/r^2$. We know of such a 
metric. It is \WHbusch. This is perhaps to be expected since the 
gauge field in 4+1 comes from the $B_{\mu\nu}$ field of 5+1 
as is the case for H-monopole solutions \banks.
Hence, there are two descriptions of the monopole 
moduli space in five dimensions: the IIA perspective where the 
metric is Taub-NUT and the M-theory perspective where 
the metric is H-monopole \WHbusch.

\newsec{Theories with an adjoint hypermultiplet}
\subsec{The metric}
It is easy to generalize the discussion above to 
$U(N_c)$ gauge theories with $N_f$ flavors and an adjoint hypermultiplet.
Instead of having the $N_c$ D2-branes end of NS 5-branes, we compactify 
the D2-branes on a circle in $x^6$ and dispense with the NS 5-branes. 
The metric that the 
D2-brane sees is the metric produced by the D4-branes
in ten dimensional space-time, which generically
goes like $1/r^3$. However, since we are taking dimension $x^6$ 
to be very small, the D4 brane metric is effectively $1/r^2$. 
This is the metric on the Coulomb branch of the two 
dimensional gauge theory 
(at least for the case of the $U(1)$ gauge field). 
The mirror theory with $U(1)^{N_f}$ with bifundamental matter fields 
is constructed by intersecting the D2-brane at points on the circle
in the $x^6$ direction by NS 5-branes.
If we consider the Higgs branch of the mirror theory, then  
the ten dimensional metric that 
D2-brane now sees is not ALE, but rather the T-dual of 
an ALE space \WHbusch\ since the 
type IIA 5-branes with one direction compact 
are T-dual to Kaluza-Klein monopoles of IIB. 
Going to M-theory decompactifies the direction transverse to 
the 5-brane of IIA and takes the corresponding metric
\WHbusch\ to the wormhole metric \metrici.

\subsec{Relations to the Matrix descriptions of the 
(1,1) and (0,2) string theories}
As was explained in \newtheories, the theories on the NS 5-branes of 
type IIA and IIB string theory decouple from the bulk in the 
limit that the string coupling constant $g_s$ goes to zero while the 
tension of the strings $M_s$ is held fixed. We will show here that
the two dimensional mirror symmetry, discussed above, relates 
the two decoupled six dimensional string theories to each other
\foot {We thank A. Hanany for suggesting this.}.
A Kaluza-Klein monopole in directions $(x^0, x^1, x^2, x^3, x^4, x^5, x^6)$ 
in M-theory with the direction $x^{10}$ compactified 
on a large circle $R_{10}$ can be described by IIA string theory with 
a D6 brane in the limit $g_s^A \rightarrow \infty$ since $g_s^A = M_sR_{10}$. 
If the 
radius of the direction $x^6$, $R_6 << 1/M_s$, then the 
theory is better described as a IIB theory with a 
D5 brane. The string coupling of the IIB theory is 
\eqn\decouple{g_s^B = {R_{10}\over R_6} \rightarrow \infty.}
S-duality takes the D5 brane to an NS 5-brane, and inverts the 
coupling constant. This is therefore the limit in which the 
IIB NS 5-brane decouples from the bulk. This is also the 
limit in which the mirror symmetry is valid.

On the other hand, the KK monopole of M-theory 
has a Matrix description in terms of D0 branes  
and KK monopoles of IIA\amigilad. We choose the direction 
$x^5$ to be the infinite boost direction. 
Therefore, here the IIA string coupling is 
$g_s^A = M_sR_5$ where $M_s^2 = M_{pl}^3 R_5$.
The coupling constant of the world volume 
theory on the D0 brane is $$g_1^2 = M_{pl}^6R_5^3.$$
Since the radius $R_6$ is small, we can T-dualize this 
into a big circle. We now have a IIB theory with a 
D1 brane in directions $(x^0, x^6)$, 
parallel to a KK monopole of IIB. T-dualizing on 
$x_{10}$, the NUT direction, we have a theory 
of D2-branes in directions $(x^0, x^6, x^{10})$ 
on a small circle intersecting 
NS 5-branes in $(x^0,x^1, x^2, x^3, x^4, x^6)$
\amigilad. 
This theory has a Higgs branch metric 
that is ALE. The coupling constant on the 
1+1 dimensional theory is 
$${1\over g_2^2} = {R_{10}R_6\over R_{5}}\tilde R_{10} = 
{R_{6}\over M_sg_s^A}$$
where $\tilde R_{10} = {1\over M_s^2R_{10}}$.
In the limit where $g_s^A \rightarrow \infty$, 
$g_2^2 \rightarrow \infty$. This is also the limit where 
\decouple\ the physics of the IIB 5-brane decouples from 
the bulk. 

The relationship between the 1+1 $U(N_c)$ gauge 
theory with $N_f$ hypermultiplets and an adjoint hypermultiplet
is more direct. 
Consider the D2-branes in $x^0, x^1, x^6$ and the 
D4 branes in $x^0, x^1, x^7, x^8, x^9$. T-dualizing along 
$x^6$ takes us to a IIB configuration with D1 branes and 
D5 branes. This is the Matrix string configuration description 
of the IIA 5-brane discussed in \ed. The limit in which 
the string coupling constant goes to zero corresponds to the 
strong coupling limit on the D2-brane. We see then that 
the two dimensional mirror symmetry between $U(N_c)$ with $N_f$ fundamentals 
and an adjoint hypermultiplet 
and $U(N_c)^{N_f}$ with bifundamental matter is a relation between the 
string theories on the NS 5-branes of IIA and IIB respectively in the 
limit that they decouple from the bulk \amigilad.

\newsec{Seiberg Duality in Two Dimensions}

In four dimensions with 4 real supercharges, there is a duality 
between an $SU(N_c)$ gauge theory with $N_f$ flavors and 
an $SU(N_f - N_c)$ gauge theory with $N_f$ flavors.
There are no adjoint matter fields in these theories, and 
the dimensions of both Higgs branch moduli spaces are the same.
This duality was realized in terms of brane constructions 
in \refs{\hw,\keg}.
Generically for theories 
with 8 supercharges the duality is spoiled by the adjoint matter fields. 
Seiberg duality in two dimensions 
could exist since the theory on the Higgs branch decouples from the 
theory on the Coulomb branch. 
One can check that $\hat c$ is the same for
the $SU(N_c)$ gauge theory with $N_f$ flavors and the 
$SU(N_f - N_c)$ gauge theory with $N_f$ flavors on the Higgs branch.
Furthermore, from the brane perspective, 
we can perform the same operations that were carried out in 
\keg\ T-dualized from four to two dimensions.
It is natural to expect that there are such dual conformal field theories 
on the Higgs branch.

\newsec{The quantum Higgs branch}
Classically a two-dimensional $U(N_c)$ gauge theory with $N_f = 1$ 
flavors and eight supercharges has no Higgs branch.
In \ed, there was speculation that quantum mechanically there 
could be such a Higgs branch. Motivation for this speculation 
came from Matrix theory where the theory of one $(0,2)$ NS 5-brane
is conjectured to be non-trivial. We have seen here that the
metric describing the dual Higgs branch should, according to
the brane picture, become deformed, due to non-trivial IR dynamics, 
into a something that has a 
Kahler potential that goes as $1/r^2$. Although we do not know exactly 
what this metric is, it seems reasonable to expect that the singularity 
in the metric for $N_f = 1$ will not be a coordinate singularity as 
it is for an ALE metric, but rather a real singularity. 
A real singularity would imply that there is a non-trivial Higgs branch 
for $N_f = 1$.

\newsec{Conclusions}

In this paper we studied the Coulomb and the Higgs branchs of 
$N = (4,4)$ theories in two dimensions.
We found using a brane construction, two different 
two dimensional theories that become equivalent in the IR.
In eleven dimensions there is a manifest symmetry 
that exchanges the Higgs branch of one theory with the Coulomb branch
of another theory.
The branes provide a way of seeing the $SU(2) \times SU(2)$ 
symmetry of the conformal field theory to which the Coulomb branch 
flows.
We have seen that ALE sigma models with $\theta = 0$ flow to 
wormhole CFTs of \callan.

In \natidia\ it was 
conjectured that the ``wormhole'' singularities of the 
Coulomb branch can be described in terms of some unspecified dual 
variables as ALE singularities 
of a dual Higgs branch. This idea was inspired by 
\ovi\ where it was conjectured
that ALE singularities in string theory are dual to 
5-branes by mirror symmetry (or rather ``fiber-wise T-duality'').
Like the mirror symmetry between Calabi-Yau manifolds, such 
conjectures are difficult to prove stemming from the 
fact that there are often no isometries to dualize upon \mp 
(although it is
possible sometimes to dualize upon broken isometries \qu ).
This is the case with the conjecture of \ovi\ that ALE singularities 
are mirror to 5-branes.
We have given here an interpretation from the brane perspective 
of this duality and explained some puzzles about the 
development of torsion on the ALE side. 
We suggest that brane configurations can be useful 
tools when analyzing sigma models with no well defined CFT.

It is interesting to note that the D-brane configuration 
that was considered in this paper, 
a D2-brane ending on an NS 5-brane, has been 
studied as a theory of self-dual strings 
in six dimensions. Little is known about self-dual strings since they
do not allow for a perturbative description.
Also, if one T-dualizes this set-up 
such that the 5-branes become ALE singularities, 
one recovers D3 branes of IIB wrapped on 
$S^2$ cycles of an ALE space another situation where 
self-dual strings arise.

Finally, 
considering all the progress that has been made in finding exact solutions of 
higher dimensional theories, one might ask why return to two-dimensions.
In this paper we analyzed certain $(4,4)$ theories via brane constructions. 
However, 
brane constructions of 2d, 3d, 4d, 5d, and 
6d with 8 
real supercharges are all related to each other by T-duality, hence 
correspondences 
between dimensions are manifest. 
A motivation for studying 2d theories is the hope that 
it will be possible also to travel up in dimension and 
relate exact solutions in 2d of interacting fixed points to higher 
dimensions where
SCFTs are known to exist, but as of yet have not been solved. Although in 2d 
there is the power of the 
two dimensional superconformal algebra at one's disposal, one can argue that 
 the branes see no difference between 2d and other dimensions 
and so one should expect 
similars solutions. 
Understanding precisely the relations between the branes and the 
superconformal 2d fixed points is essential for such a program.

\newsec{Acknowledgements}
I would like to thank P. Aspinwall, A. Hanany, K. Intriligator, S. Ramgoolam, 
E. Sharpe, N. Seiberg, S. Sethi, and E. Witten for helpful discussions. 
This work was supported 
in part by DOE grant number DE-FG02-91ER40671.

\listrefs

\end